\documentclass[letterpaper]{jpconf}
\usepackage{epsfig,bm,feynmf,amsmath,amssymb}
\usepackage{graphics}
\usepackage[normalem]{ulem}  
\usepackage[dvips]{color} 
\newcommand{\com}[1]{{\sf\color[rgb]{0,0,1}{#1}}}

\renewcommand\sout{\bgroup \color{red} \ULdepth=-.5ex \ULset}

\begin{document}

\title{Jet Fragmentation via Recombination of Parton Showers}

\author{Kyong Chol Han$^1$, Rainer J Fries$^1$, and Che Ming Ko$^1$}

\address{$^1$ Cyclotron Institute and Department of Physics and Astronomy, Texas A$\&$M University, College Station, TX 77843-3366, USA}

\ead{khan@comp.tamu.edu; rjfries@comp.tamu.edu; ko@comp.tamu.edu}

\begin{abstract}
We study hadron production in jets by applying quark recombination to 
jet shower partons. With the jet showers obtained from PYTHIA and
augmented by additional non-perturbative effects, we compute hadron spectra in $e^{+}+e^-$ collisions at $\sqrt{s}=200~\rm{GeV}$. Including contributions from resonance decays, 
we find that the resulting transverse momentum spectra for 
pions, kaons, and protons reproduce reasonably those from the
string fragmentation as implemented in PYTHIA.
\end{abstract}

\section{Introduction}
Hadron production from energetic jets in high-energy nuclear collisions 
is often parameterized through fragmentation functions using the 
universality of the process as given by QCD factorization theorems \cite{Collins:1981uw}.
On a more microscopic level, one can achieve a good description of jet hadronization through a perturbative evolution of the parton spectrum inside the jet using DGLAP splitting kernels from the largest momentum scale 
to some lower cutoff $Q_0$, followed by a non-perturbative 
hadronization model like the Lund string model applied to the parton shower. The event generator 
PYTHIA \cite{Sjostrand:2006za} has succes\com{s}fully implemented such a 
strategy to describe hadron production in $e^{+}+e^-$ collisions at 
high energies and in numerous other processes.

On the other hand, it has been suggested that hadron production can often be
described through a process of quark recombination or coalescence 
\cite{Das:1977cp,Fries:2003vb,Greco:2003xt,Greco:2003mm,Fries:2003kq}.
It is an intriguing idea to combine the concepts of quark recombination and
parton showers since it would easily generalize to the hadronization of
jets in dense environments as in relativistic heavy ion collisions.
Hwa and Yang have pioneered the application of quark recombination to 
jet showers in Ref.\ \cite{Hwa:2003ic}. However, the parton showers in their work were not 
obtained from first principles but fitted to measured hadron spectra. They
also ignored the smallness of the number of partons in each jet 
by applying recombination to averaged spectra instead of event-by-event
showers. Fluctuations from shower to shower are expected to lead to significant
changes to the hadron spectrum at large hadron momentum fraction $z$.

Using $e^{+}+ e^{-}$ collisions as an example, we show that vacuum fragmentation functions can 
be reproduced in an event-by-event approach that uses a perturbative 
shower evolution in the vacuum down to a scale $Q_0$, a minimal modeling 
of non-perturbative effects on partons below $Q_0$, and finally 
recombination of the quarks and antiquarks into hadrons.

\section{Shower Partons from Electron-Positron Annihilation in PYTHIA}

\begin{figure}[h]
\begin{center}
\includegraphics[width=1.0\textwidth]{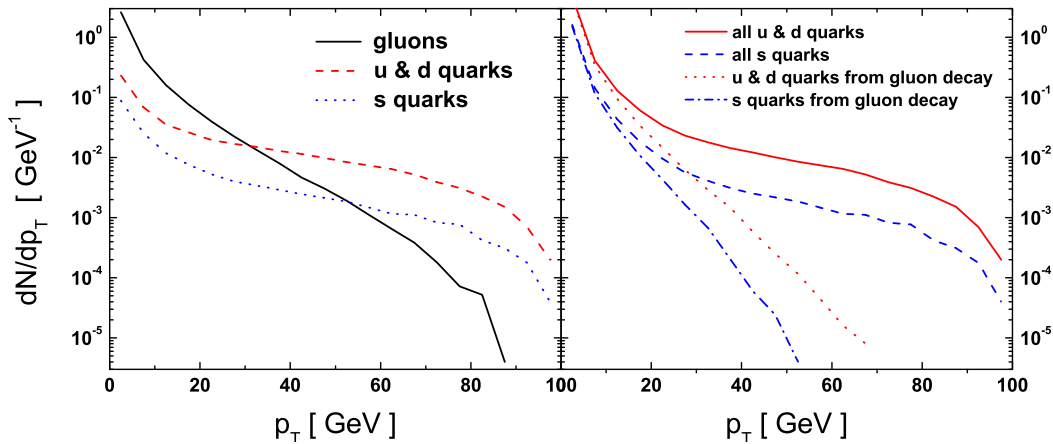}
\caption{(Color online) Transverse momentum spectra of shower partons before (left panel) and after (right panel) including gluon decays into quark-antiquark pairs.}
\label{gluondecaygraph}
\end{center}
\end{figure}

Using PYTHIA 6.3 \cite{Sjostrand:2006za}, we generate an ensemble of parton showers from $e^{+}+e^{-}$ collisions at center-of-mass energy $\sqrt{s}=200~\rm{GeV}$. The light quark, strange quark and gluon spectra as functions of momentum at a cutoff scale $Q_0=1~{\rm GeV}$ are shown in the left panel of Fig.~\ref{gluondecaygraph}. Since gluons in the parton showers cannot directly recombine into hadrons, we force them to decay into quark-antiquark pairs after the perturbative evolution is stopped. This is a non-perturbative process, and we use constituent quark masses of $m_{u,d}=0.3~\rm{GeV}$ for light quarks and $m_s=0.45~\rm{GeV}$ for strange quarks with the remaining gluon virtuality randomly chosen in the interval between $2 m_{u,d}$ and $Q_0$. The transverse momentum spectra of the decay light and strange quarks together with the total light and strange quark spectra are shown in the right panel of Fig.\ \ref{gluondecaygraph}.

\section{Quark Recombination}

Following the approach to quark recombination in Ref.~\cite{Greco:2003xt} the probability for quarks and antiquarks to form a hadron is taken to be proportional to the quark Wigner function of the hadron. Assuming the quark wave functions of the hadrons to be those of harmonic oscillators, the hadron Wigner function is then Gaussian in both the coordinate and the momentum space. Here we neglect spatial information, assuming shower partons with similar momenta to be close in space. Hence we only consider the momentum-space Wigner function. 

For the production of a meson from a quark and an antiquark, the probabilities are
\begin{align}
\label{WigM}
f_{s}({\bf k}) = 8g_{\rm{M}}\exp \left(-\frac{{\bf k}^2}{\Delta^2}\right)~~{\rm and}~~
f_{p}({\bf k}) =g_{M}\left(-8+\frac{16}{3}\frac{{\bf k}^2}{\Delta^2}\right)\exp\left(-\frac{{\bf k}^2}{\Delta^2}\right),
\end{align}
respectively\com{,} for $s$-wave and $p$-wave mesons \cite{Cho:2011ew}. Here $\bf k$ is the relative momentum between the quark and antiquark in the meson rest frame, $\Delta$ is the width of the quark and antiquark relative momentum distribution in the meson, and $g_{M}$ is the statistical factor for the colored, spin-$1/2$ quark and antiquark to form a colorless meson. For the mesons $\pi$, $\rho$, $a_1$, $\omega$, $f_2(1270)$, $f_1(1285)$, $K^{*}$, $K_{1}(1270)$, and $K_1^*(1410)$ considered in our study, we have $g_{\pi}=g_{K}=1/36$, $g_{\rho}=g_\omega=g_{a_1}=g_{f_1}=g_{K^*}=g_{K_1}=g_{K^*_1}=1/12$, and $g_{f_2}=5/36$. Because of the color charge of a gluon, its decay quark and antiquark are excluded from recombining into a color singlet meson.

For $s$-wave baryons, the production probability from three quarks is \cite{Oh:2009zj}
\begin{eqnarray}
\label{WigB}
f_{B}({\bf k}_1,{\bf k}_2) = 64 g_{B} \exp\Big(-\frac{{\bf k}_1^{2}}{\Delta^2}-\frac{3{\bf k}_2^{2}}{4\Delta^2}\Big),
\end{eqnarray}
where
${\bf k}_1=({\bf p}_1^\prime-{\bf p}_2^\prime)/2$ and ${\bf k}_2=({\bf p}_1^\prime+{\bf p}_2^\prime-2{\bf p}_3^\prime)/3$
are the relative momenta of the recombining quarks of momenta ${\bf p}_1^\prime$, 
${\bf p}_2^\prime$, and ${\bf p}_3^\prime$ in the rest frame of the baryon. The statistical factors for the 
baryons $N$, $N_{1}(1440)$, $\Delta$, and $\Delta_1(1600)$ considered in our study are $g_{N}=g_
{N_1}=1/108$ and $g_{\Delta}=g_{\Delta_1}=1/54$.

\section{Pion, Kaon and Nucleon Transverse Momentum Spectra}

\begin{figure}[h]
\begin{center}
\includegraphics[width=1\textwidth]{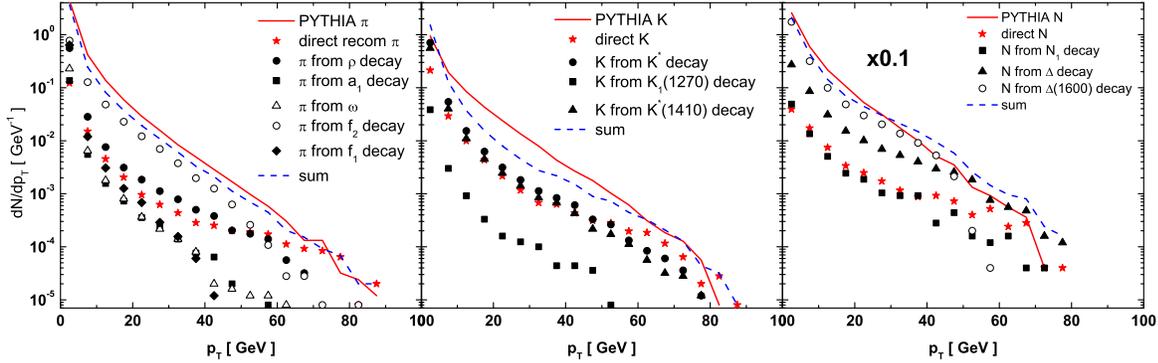}
\caption{(Color online) Transverse momentum spectra of pions (left panel), kaons (middle panel), and nucleons (right panel).}
\label{recomb}
\end{center}
\end{figure}

The momentum width $\Delta$ is related to the oscillator constant in the quark wave function. Here we take them as 
parameters. Using $\Delta_{\pi}=\Delta_{\rho}=\Delta_{a_1}=\Delta_\omega=\Delta_{f_2}=\Delta_{f_1}=0.43~\rm{GeV}$, 
$\Delta_{K}=\Delta_{K^*}=\Delta_{K_1}=\Delta_{K_1^*}=0.58~\rm{GeV}$, and $\Delta_{N}=\Delta_{N_1}=\Delta_{\Delta_1}=0.62~{\rm GeV}$, we have calculated their respective transverse momentum spectra from the recombination of quarks and antiquarks on an event-by-event basis. The spectra of directly produced pions, kaons and nucleons are shown by star symbols in the left, middle, and right panels of Fig.~\ref{recomb}, respectively. They underestimate those obtained from PYTHIA via string fragmentation shown by solid lines. After including contributions from the decay of resonances via
\begin{eqnarray}
&& \rho \rightarrow \pi\pi,~~~~a_{1} \rightarrow \rho\pi \rightarrow \pi\pi\pi,~~~~\omega\to\pi\pi\pi,
~~~~f_2\to\pi\pi,~~~~f_1\to 4\pi,\nonumber\\
&& K^{*} \rightarrow K\pi,~~~~K_{1}\rightarrow K\rho \rightarrow K\pi\pi,~~~~K_1^*\to K^*\pi\to K\pi\pi,\nonumber\\
&&N_{1}\rightarrow N\pi(65\%)~~{\rm or}~~N\rho(35\%)\rightarrow N\pi\pi, \nonumber\\
&& \Delta \rightarrow N\pi,~~~\Delta_{1}\rightarrow N\pi, \nonumber
\end{eqnarray}
we significantly improve the comparison with the PYTHIA results as shown by the dashed lines in Fig.~\ref{recomb}. 

\begin{figure}[h]
\begin{center}
\includegraphics[width=0.6\textwidth]{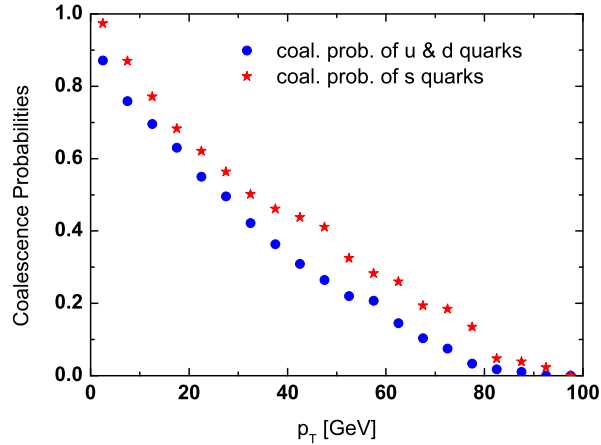}
\caption{(Color online) Probabilities of light and strange quarks to recombine into hadrons as functions of their transverse momenta.}
\label{probremn}
\end{center}
\end{figure}

We note that not all quarks and antiquarks are used in the recombination model to form hadrons. As shown in Fig.~\ref{probremn}, the recombination probability is close to one for low momentum quarks and decreases with increasing momentum. The small number of non-recombined quarks thus consists mostly of leading quarks that are far removed in momentum space from the rest of the parton shower. Their contribution to hadron production can be included via independent fragmentations at the low momentum scale $Q_0$. Because of their small numbers, the final hadron spectra are, however, not much affected.

\section{Conclusions}

Our study indicates that hadron production from jets can be described by the recombination of shower partons inside the jets. Our approach provides a promising framework to study the medium modification of fragmentation functions in the presence of a quark gluon plasma (QGP) by allowing shower partons to recombine with the thermal partons in the QGP.

\section*{Acknowledgments}

This work was supported in part by the U.S. National Science Foundation under Grants Nos.\ PHY-0847538 and PHY-1068572, the JET Collaboration and the US Department of Energy under Contract No.\ DE-FG02-10ER41682, and the Welch Foundation under Grant No.\ A-1358.

\end{document}